\documentclass[conference]{IEEEtran}
\usepackage{enumitem}
\usepackage{graphicx,multicol} 
\usepackage{epsfig} 
\usepackage{amsmath,amsthm} 
\usepackage{amssymb}  
\usepackage{float}
\usepackage{cases,setspace,adjustbox}
\usepackage{cite}
\usepackage{algorithmic}
\usepackage[ruled,vlined]{algorithm2e}
\usepackage{array}
\usepackage[update,prepend]{epstopdf}
\usepackage{eqparbox}
\usepackage{subfig}
\usepackage{fixltx2e}
\usepackage{url}
\usepackage{amsthm}
\usepackage{array}
\usepackage{caption}
\usepackage{xcolor}
\usepackage{mathtools}
\usepackage{bbm}
\usepackage{multirow}
\newcommand{\vect}[1]{\boldsymbol{#1}}
\usepackage{booktabs}
\newcommand{\ra}[1]{\renewcommand{\arraystretch}{#1}}

\newcommand{\pidle}{p_{\text{I}}}
\newcommand{\psucci}[1]{p_{\text{S}}^{(#1)}}
\newcommand{\pbusy}[1]{p_{\text{S}}^{(#1)}}
\newcommand{\pcol}{p_{\text{C}}}
\newcommand{\psucc}{p_{\text{S}}}

\newcommand{\lidle}{\sigma_\text{I}}
\newcommand{\lsucc}{\sigma_\text{S}}
\newcommand{\lcol}{\sigma_\text{C}}

\newcommand{\AoI}[1]{\Delta_{#1}}		
\newcommand{\initAoIT}[2]{\delta^{-}_{#1{#2}}} 
\newcommand{\AoIT}[2]{{\Delta}_{#1{#2}}}
\newcommand{\AvgAoIT}[2]{\widetilde{\Delta}_{#1{#2}}}
\newcommand{\AvginitAoIT}[2]{\widetilde{\Delta}^{-}_{#1{#2}}}

\newcommand{\Thr}[1]{\Gamma_{#1}}
\newcommand{\AvgThr}[1]{\widetilde{\Gamma}_{#1}}

\newcommand{\tauD}{\tau_{\mathrm{A}}}
\newcommand{\tauW}{\tau_{\mathrm{T}}}
\newcommand{\AO}{\textrm{AON}}
\newcommand{\TO}{\textrm{TON}}
\newcommand{\uD}{u_{\mathrm{A}}}
\newcommand{\uW}{u_{\mathrm{T}}}
\newcommand{\ND}{N_{\mathrm{A}}}
\newcommand{\NW}{N_{\mathrm{T}}}

\newcommand{\T}{\mathcal{T}}
\newcommand{\I}{\mathcal{I}}

\newtheorem{proposition}{Proposition}
\newcommand{\SG}[1]{{\color{black} {#1}}}

\begin{document}
\renewcommand{\arraystretch}{1.15}
\renewcommand{\vec}[1]{\mathbf{#1}}
\title{Coexistence of Age and Throughput Optimizing Networks: A Game Theoretic Approach}
\author{Sneihil Gopal$^{*}$, Sanjit K. Kaul$^{*}$ and Rakesh Chaturvedi$^{\dagger}$\\
$^{*}$Wireless Systems Lab, IIIT-Delhi, India \\
$^{\dagger}$Department of Social Sciences \& Humanities, IIIT-Delhi, India\\
\{sneihilg, skkaul, rakesh\}@iiitd.ac.in}
\maketitle
\begin{abstract}
Real-time monitoring applications have Internet-of-Things (IoT) devices sense and communicate information (status updates) to a monitoring facility. Such applications desire the status updates available at the monitor to be fresh and would like to minimize the age of delivered updates. Networks of such devices may share wireless spectrum with WiFi networks. Often, they use a CSMA/CA based medium access similar to WiFi. However, unlike them, a WiFi network would like to provide high throughputs for its users.

We model the coexistence of such networks as a repeated game with two players, an age optimizing network ($\AO$) and a throughput optimizing network ($\TO$), where an $\AO$ aims to minimize the age of updates and a $\TO$ seeks to maximize throughput. We define the stage game, parameterized by the average age of the $\AO$ at the beginning of the stage, and derive its mixed strategy Nash equilibrium (MSNE). We study the evolution of the equilibrium strategies over time, when players play the MSNE in each stage, and the resulting average discounted payoffs of the networks. It turns out that it is more favorable for a $\TO$ to share spectrum with an $\AO$ in comparison to sharing with another $\TO$. The key to this lies in the MSNE strategy of the $\AO$ that occasionally refrains all its nodes from transmitting during a stage. Such stages allow the $\TO$ competition free access to the medium.
\end{abstract}
\section{Introduction}
\label{sec:intro}
The ubiquity of Internet-of-Things (IoT) devices has led to the emergence of applications that require these devices to sense and communicate information (status updates) to a monitoring facility, or share with other devices, in a timely manner. These applications include real-time monitoring systems such as disaster management, environmental monitoring and surveillance~\cite[references therein]{fanet}, which require timely-delivery of information updates to a common ground station for better system performance, to networked control systems like vehicular networks, where each vehicle broadcasts status (position, velocity, steering angle, and etc.) to nearby vehicles in real-time for safety and collision avoidance\cite{vanet}. 

Such networks often share the wireless spectrum with WiFi networks. For instance, the Federal Communications Commission (FCC) in the US opened up the $5.85-5.925$ GHz band, previously reserved for vehicular communication, for use by high throughput WiFi ($802.11$ ac) devices, leading to the possibility of coexistence between WiFi and vehicular networks~\cite{liu2017}. Similarly, IoT devices like Unmanned Aerial Vehicles (UAVs), equipped with $802.11$ a/b/g/n technology, operate in the $2.4$ and $5$ GHz bands in use by WiFi networks.

While a network of IoT devices would like to optimize freshness of status updates, a WiFi network would like to provide high throughputs for its users. We quantify freshness using the metric of age of information~\cite{kaul2011minimizing} and refer to the former network as an age optimizing network ($\AO$) and to the latter as a throughput optimizing network ($\TO$).

In this work, we investigate the coexistence of an $\AO$ and a $\TO$ when both networks use a WiFi like CSMA/CA based medium access from a MAC layer perspective. We use a repeated game theoretic approach. We assume networks are selfish players and aim to optimize their respective utilities i.e. an $\AO$ aims to minimize the age of updates and a $\TO$ seek to maximize its throughput. Our specific contributions include
\begin{itemize}
\item We model the interaction between an $\AO$ and a $\TO$ in each CSMA/CA slot as a non-cooperative stage game, define the stage game which is parameterized by the average age of the $\AO$ at the beginning of the stage and derive its mixed strategy Nash equilibrium (MSNE). 
\item Our analysis \SG{shows} that the equilibrium strategy of each network is independent of the other network 
\SG{and the equilibrium strategy of the $\AO$ in each stage is a function of the average age seen at the beginning of the stage.} We study the subgame perfect equilibria that involves players playing the equilibrium strategy in each stage and analyse the evolution of these strategies over time. 
\item We show that unlike prior works on coexistence of CSMA/CA based networks, where networks access the medium aggressively to maximize their respective utilities~\cite{mario2005}, in $\AO$-$\TO$ coexistence, the requirement of timely updates~\cite{kaul2011minimizing} by the $\AO$ makes it conservative. Consequently, spectrum sharing with an $\AO$ becomes beneficial for a $\TO$ in contrast to sharing with another $\TO$. Specifically, we show that the equilibrium strategy of the $\AO$ occasionally refrains the $\AO$ nodes from transmitting during a stage in order to ensure freshness of updates. Such stages allow the $\TO$ competition free access to the medium, therefore, improving its payoff. 
\end{itemize}

The rest of the paper is organized as follows. Section~\ref{sec:related} describes the related works. The network model is described in Section~\ref{sec:model}. This is followed by formulation of the game in Section~\ref{sec:game}. In Section~\ref{sec:results}, we discuss the results and we conclude with a summary of our observations in Section~\ref{sec:conclusion}.
\section{Related Work}
\label{sec:related}
Works such as~\cite{liu2017} and~\cite{naik2017} study the impact of vehicular communications on WiFi and vice versa. In these works, authors look at the coexistence of vehicular and WiFi networks as the coexistence of two CSMA/CA based networks, where the packets of vehicular network take precedence over that of WiFi. In contrast to~\cite{liu2017} and~\cite{naik2017}, we look at the coexistence problem as that of coexistence of networks which have equal access rights to the spectrum, use similar access mechanisms but have different objectives.
 
In~\cite{UAV_delay} and~\cite{UAV_throughput_2} authors consider UAV applications. While in~\cite{UAV_delay} authors derive an optimum strategy for timely delivery of data so as to minimize communication delay, in~\cite{UAV_throughput_2} authors evaluate $802.11$ n and $802.11$ ac in a UAV setting in terms of achievable throughput. Delay and throughput used in the aforementioned works are commonly used performance metrics, however, they fail to measure the freshness of the updates. 
In contrast to~\cite{UAV_delay} and~\cite{UAV_throughput_2}, we employ the age of information metric, which adequately captures the freshness of updates. Works such as~\cite{sun2017update} and~\cite{yates2015lazy} investigate age of information in wireless networks. 

Note that while throughput as the payoff function has been extensively studied from the game theoretic point of view (for example, see~\cite{mario2005,chen2010}), age as a payoff function has not garnered much attention yet. In~\cite{impact2017} and~\cite{YinAoI2018}, authors study an adversarial setting where one player aims to maintain the freshness of information updates while the other player aims to prevent this. Also, in our preliminary work~\cite{SGAoI2018}, we propose a game theoretic approach to study the coexistence of DSRC (Dedicated Short Range communication aka vehicular communications) and WiFi, where the DSRC network desires to minimize the time-average age of information and the WiFi network aims to maximize the average throughput. We studied the one-shot game and evaluated the Nash and Stackelberg equilibrium strategies. However, the model in~\cite{SGAoI2018} did not capture well the interaction of networks, evolution of their respective strategies and payoffs over time, which the repeated game model allows us to capture in this work. In this work, via the repeated game model we are able to shed better light on the $\AO$-$\TO$ interaction and how their different utilities distinguish their coexistence from the coexistence of two utility maximizing CSMA/CA based networks. 


\section{Network Model}
\label{sec:model}
We consider a network which consists of $\ND$ age optimizing and $\NW$ throughput optimizing nodes that contend for access to the shared wireless medium.~\SG{We assume the number of nodes is time-invariant.} In general, both $\AO$ and $\TO$ nodes access the medium using a CSMA/CA based mechanism in which nodes use contention windows (CW) and one or more backoff stages to gain access to the medium\footnote{In CSMA/CA, nodes employ a window based backoff mechanism to gain access to the medium. The node first senses the medium and if the medium is busy it chooses a backoff time uniformly from the interval $[0, w-1]$, where $w$ is set equal to $CW_{min}$. The interval is doubled after each unsuccessful transmission until the value equals $CW_{max} = 2^{m}CW_{min}$, where $m$ is the maximum backoff stage.}. 
We model this mechanism as in~\cite{bianchi}. 

We assume that all nodes can sense each other's packet transmissions. This allows modeling the CSMA/CA mechanism as a slotted multiaccess system. A slot may be an idle slot in which no node transmits a packet or it may be a slot that sees a successful transmission. This happens when exactly one node transmits. If more than one node transmits, none of the transmissions are successfully decoded and the slot sees a collision. Further, we assume that all nodes always have a packet to send. The modeling in~\cite{bianchi} shows that the CSMA/CA settings of minimum contention window ($CW_{min}$), number of backoff stages and the number of nodes can be mapped to the probability with which a node attempts transmission in a slot. We use this probability to calculate the probabilities defined next. We will define the probabilities of interest for a certain network of nodes indexed $\{1,2,\dots,N\}$.

Let $\tau_i$ denote the probability with which node $i$ attempts transmission in a slot. Let $\pidle$ be the probability of an idle slot, which is a slot in which no node transmits. We have
\begin{align}
\pidle = \prod\limits_{i=1}^{N} (1-\tau_i).
\label{Eq:probidle}
\end{align}
Let $\psucci{i}$ be the probability of a successful transmission by node $i$ in a slot and let $\psucc$ be the probability of a successful transmission in a slot. We say that node $i$ sees a busy slot if in the slot node $i$ doesn't transmit and exactly one other node transmits. Let $\pbusy{-i}$ be the probability that a busy slot is seen by node $i$. Let $\pcol$ be the probability that a collision occurs in a slot. We have
\begin{align}
\psucci{i}&= \tau_i\prod\limits_{\substack{j=1\\j\neq i}}^{N} (1-\tau_j),\quad \psucc= \sum\limits_{i=1}^{N}\psucci{i},\nonumber\\
\psucci{-i}&= \sum\limits_{\substack{j=1\\j\neq i}}^{N}\tau_j\prod\limits_{\substack{k=1\\k\neq j}}^{N}(1-\tau_k)\text{ and } \pcol = 1-\pidle-\psucc.
\label{Eq:probsucc}
\end{align}

Let $\sigma_I, \sigma_S$ and $\sigma_C$ denote the lengths of an idle, successful, and collision slot, respectively. In this work, we assume $\lsucc = \lcol$. The other case of practical interest, for when using RTS/CTS, is $\lsucc > \lcol$. The analysis for this case can be carried out in a similar manner to that in this paper. We skip the details in this paper.

Next we define the throughput of a $\TO$ node and the age of an $\AO$ node in terms of the above probabilities and slot lengths. 

\subsection{Throughput of a $\TO$ node over a slot}
Let the rate of transmission be fixed to $r$ bits/sec in any slot. Define the throughput $\Thr{i}$ of a $\TO$ node $i\in \{1,2,\dots,\NW\}$ in a slot as the number of bits transmitted successfully in the slot. This is a random variable with probability mass function (PMF)
\begin{align}
P[\Thr{i} = \gamma] =
    \begin{cases}
      \psucci{i} &\gamma = \lsucc r, \\
      1-\psucci{i} &\gamma = 0,\\
      0 & \text{otherwise.}
    \end{cases}
\label{Eq:ThrPMF}
\end{align}
Using~(\ref{Eq:ThrPMF}), we define the average throughput $\AvgThr{i}$ of node $i$ as
\begin{align}
\AvgThr{i} = \psucci{i}\lsucc r.
\label{Eq:Thr}
\end{align}
The average throughput of $\TO$ in a slot is
\begin{align}
\AvgThr{} = \frac{1}{\NW}\sum\limits_{i=1}^{\NW}  \AvgThr{i}.
\label{Eq:NetThr}
\end{align}
Note that the throughput in a slot is independent of that in the previous slots.

\subsection{Age of an $\AO$ node over a slot}
\begin{figure}[t]
\begin{center}
\includegraphics[width=\columnwidth]{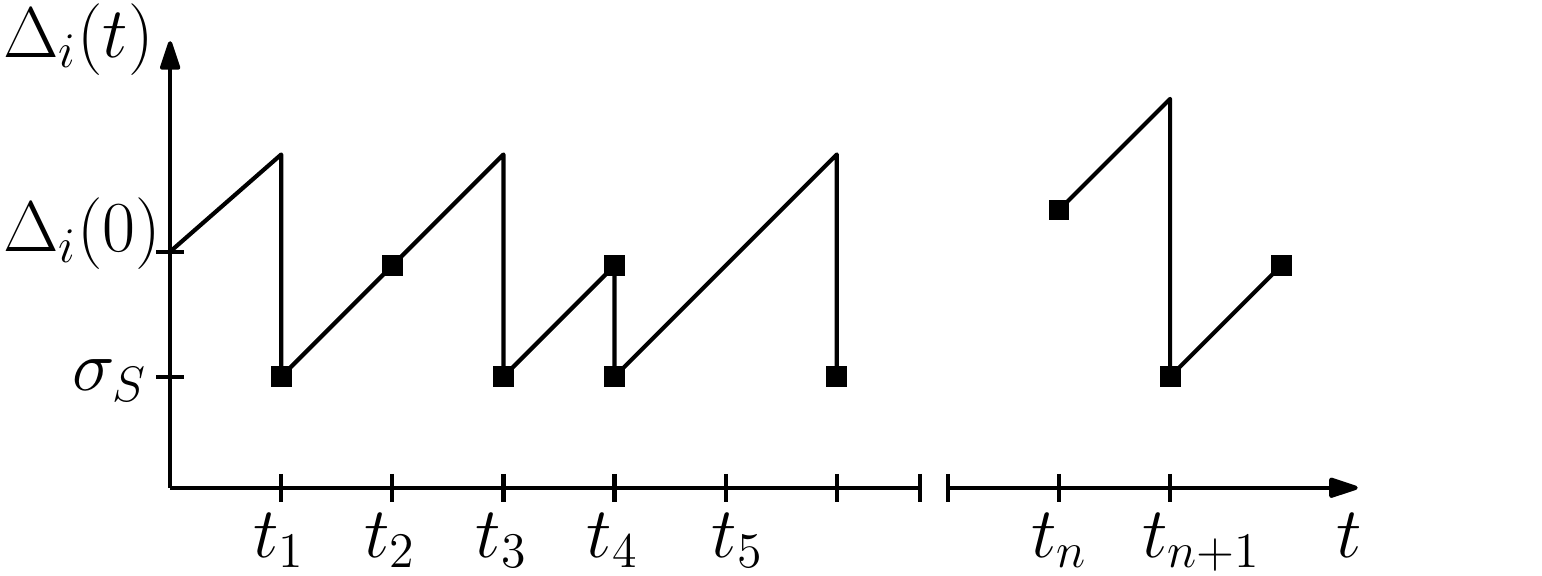}
\end{center}
\caption{\small $\AO$ node $i$'s sample path of age $\AoI{i}(t_{n})$. $\AoI{i}(0)$ is the initial age. A successful transmission resets the age to $\lsucc$. The time instants $t_{n}$, where, $n \in \{1,2,\dots\}$, show the slot boundaries. The inter-slot intervals are determined by the type of slot.}
\label{fig:instaAoI}%
\end{figure}

Let $\Delta_{i}(t)$ be the status update age of an $\AO$ node $i \in \{1,2,\dots,\ND\}$ at other nodes in the AON at time $t$. \SG{When the freshest update of node $i$ at $\AO$ node $j \in \{1,2\dots,\ND\}\setminus i$ at time $t$ is time-stamped $u(t)$, the \textit{status update age}, or simply the \textit{age}, of node $i$ at node $j$ is defined as $\Delta_{i}(t) = t - u(t)$.}
We assume that a status update packet that $\AO$ node $i$ attempts to transmit in a slot contains an update that is fresh at the beginning of the slot. 
As a result, node $i$'s age either resets to $\lsucc$ if a successful transmission slot occurs 
or increases by $\lidle$, $\lcol$ or $\lsucc$ at all other nodes in the $\AO$, respectively, in case an idle slot, collision slot or a busy slot occurs. 
Note that node $i$'s age at the end of a slot is determined by its age at the beginning of the slot and the type of the slot. Figure~\ref{fig:instaAoI} shows an example sample path of the age $\AoI{i}(t)$ of a certain $\AO$ node $i\in\{1,2,\dots,\ND\}$. In what follows we will drop the explicit mention of time $t$ and let $\AoIT{i}{}$ be the age observed at the end and $\Delta^{-}_{i}$ be the age observed at the beginning of a given slot by node $i$. 

The age $\AoIT{i}{}$, observed by node $i$ at the end of a slot is thus a random variable with PMF conditioned on age at the beginning of a slot given by
\begin{align}
P[\AoIT{i}{} = \delta_{i}|\Delta^{-}_{i} = \initAoIT{i}{}] = 
	\begin{cases}
	\pidle & \delta_{i} = \initAoIT{i}{}+\lidle ,\\
	\pcol & \delta_{i} = \initAoIT{i}{}+\lcol,\\
	\psucci{-i} & \delta_{i} = \initAoIT{i}{}+\lsucc,\\
	\psucci{i} & \delta_{i} = \lsucc,\\
	0 & \text{otherwise.}
	\end{cases}
	\label{Eq:AoIPMF}
\end{align}
Using~(\ref{Eq:AoIPMF}), we define the conditional expected age
\begin{align}
\AvgAoIT{i}{} &\overset{\Delta}{=} E[\AoIT{i}{} = \delta_{i}|\Delta^{-}_{i} = \initAoIT{i}{}].\nonumber\\
&=(1-\psucci{i})\initAoIT{i}{}+(\pidle\lidle+\psucc\lsucc+\pcol\lcol).
\label{Eq:AoI}
\end{align}
The average age of an $\AO$ at the end of the slot is
\begin{align}
\AvgAoIT{}{}& = \frac{1}{\ND}\sum\limits_{i=1}^{\ND}\AvgAoIT{i}{}.
\label{Eq:NetAoI}
\end{align}
\vspace{-2em}
\section{The Age-Throughput Optimizing Repeated Game}
\label{sec:game}
We define a repeated game to model the interaction between an $\AO$ and a $\TO$. In every CSMA/CA slot, networks must compete for access with the goal of maximizing their expected payoff over an infinite horizon (a countably infinite number of slots). We capture the interaction in a slot as a non-cooperative stage game $G$. The interaction over the infinite horizon is modeled as the stage game played repeatedly in every slot and defined as $G^{\infty}$. Next, we define these games in detail.
\subsection{Stage game}
\label{sec:one-shot}
We define a parameterized strategic one-shot game $G = (\mathcal{N},(\mathcal{S}_k)_{k\in\mathcal{N}},(u_k)_{k\in\mathcal{N}},\AvginitAoIT{}{})$, where $\mathcal{N}$ is the set of players, $\mathcal{S}_k$ is the set of strategy of player $k$, $u_k$ is the payoff of player $k$ and $\AvginitAoIT{}{}$ is the additional parameter input to the game $G$, which is the average age~(\ref{Eq:NetAoI}) of the $\AO$ seen at the beginning of the slot. We define the game $G$ in detail. 
\begin{itemize}
\item \textbf{Players:} We have two players namely $\AO$ ($\mathrm{A}$) and $\TO$ ($\mathrm{T}$). Specifically, $\mathcal{N} = \{\mathrm{A},\mathrm{T}\}$.
\item \textbf{Strategy:} Let $\T$ denote transmit and $\I$ denote idle. For $\AO$ comprising of $\ND$ nodes, the set of pure strategies is $\mathcal{S}_{A} \triangleq \mathbb{S}_{1}\times \mathbb{S}_{2} \times \dots \times \mathbb{S}_{\ND}$, where $\mathbb{S}_{i} = \{\T,\I\}$ is the set of strategies for a certain $\AO$ node $i\in\{1,2,\dots,\ND\}$. Similarly, for the $\TO$ comprising of $\NW$ nodes, the set of pure strategies is $\mathcal{S}_{\mathrm{T}} \triangleq \mathbb{S}_{1}\times \mathbb{S}_{2} \times \dots \times \mathbb{S}_{\NW}$, where $\mathbb{S}_{i} = \{\T,\I\}$ is the set of strategies for a certain $\TO$ node $i\in\{1,2,\dots,\NW\}$. 

We allow networks to play mixed strategies. In general, for the strategic game $G$, we can define $\boldsymbol{\Phi}_{k}$ as the set of all probability distributions over the set of strategies $\mathcal{S}_{k}$ of player $k$, where $k\in \mathcal{N}$. A mixed strategy for player $k$ is an element $\phi_{k} \in \boldsymbol{\Phi}_{k}$, such that $\phi_{k}$ is a probability distribution over $\mathcal{S}_{k}$. For example, for an $\AO$ with $\ND = 2$, the set of pure strategies is $\mathcal{S}_{A} = \mathbb{S}_{1} \times \mathbb{S}_{2} = \{(\T,\T),(\T,\I),(\I,\T),(\I,\I)\}$ and the probability distribution over $\mathcal{S}_{A}$ is $\phi_{A}$ such that $\phi_{A}(s_{A})\geq 0$ for all $s_A \in \mathcal{S}_{A}$ and $\sum_{s_A \in \mathcal{S}_{A}} \phi_{A}(s_{A}) = 1$.  

In this work, we restrict ourselves to the space of probability distributions such that the mixed strategies of an $\AO$ are a function of $\tauD$ and that of a $\TO$ are a function of $\tauW$, where $\tauD$ and $\tauW$, as defined earlier, are the probabilities with which nodes in an $\AO$ and a $\TO$, respectively, attempt transmission in a slot. Specifically, we force all the nodes to choose the same probability to attempt transmission. As a result, the probability distribution $\phi_{A}$ for an $\AO$ with $\ND = 2$, parameterized by $\tauD$, is $\phi_{A} = \{\phi_{A}(\T,\T), \phi_{A}(\T,\I), \phi_{A}(\I,\T), \phi_{A}(\I,\I)\} = \{\tauD^2, \tauD(1-\tauD), (1-\tauD)\tauD, (1-\tauD)^{2}\}$.

Similarly, the probability distribution $\phi_{W}$ for a $\TO$ with $\NW = 2$, parameterized by $\tauW$, is $\phi_{W} = \{\phi_{W}(\T,\T), \phi_{W}(\T,\I), \phi_{W}(\I,\T), \phi_{W}(\I,\I)\} = \{\tauW^2, \tauW(1-\tauW), (1-\tauW)\tauW, (1-\tauW)^{2}\}$. 

We therefore allow the $\AO$ and the $\TO$ to choose $\tauD \in [0,1]$ and $\tauW \in [0,1]$, respectively, to compute the mixed strategies.

\item \textbf{Payoffs:} We have $\NW$ throughput optimizing nodes that attempt transmission with probablity $\tauW$ and $\ND$ age optimizing nodes that attempt transmission with probability $\tauD$. Thus, by substituting $\tau_{i} = \tauW$ for $i$ that is a $\TO$ node and $\tau_{i} = \tauD$ for $i$ that is an $\AO$ node, we can calculate the probabilities~(\ref{Eq:probidle})-(\ref{Eq:probsucc}) of both the $\TO$ and the $\AO$ nodes. The probabilities can be substituted in~(\ref{Eq:ThrPMF})-(\ref{Eq:Thr}) and~(\ref{Eq:AoIPMF})-(\ref{Eq:AoI}), respectively, to obtain the average throughput in~(\ref{Eq:NetThr}) and average age in~(\ref{Eq:NetAoI}). We use these to obtain the stage payoffs $\uW$ and $\uD$ of the $\TO$ and the $\AO$, respectively. They are
\begin{align}
\uW(\tauD,\tauW) &= \AvgThr{}{}(\tauD,\tauW)\label{Eq:thrpayoff},\\
\uD(\tauD,\tauW) &= -\AvgAoIT{}{}(\tauD,\tauW)\label{Eq:agepayoff}.
\end{align}
The networks would like to maximize their payoffs.
\end{itemize}

\subsection{Mixed Strategy Nash Equilibrium}
\begin{figure}[t]
  \centering
  \subfloat[]{\includegraphics[width=0.49\columnwidth]{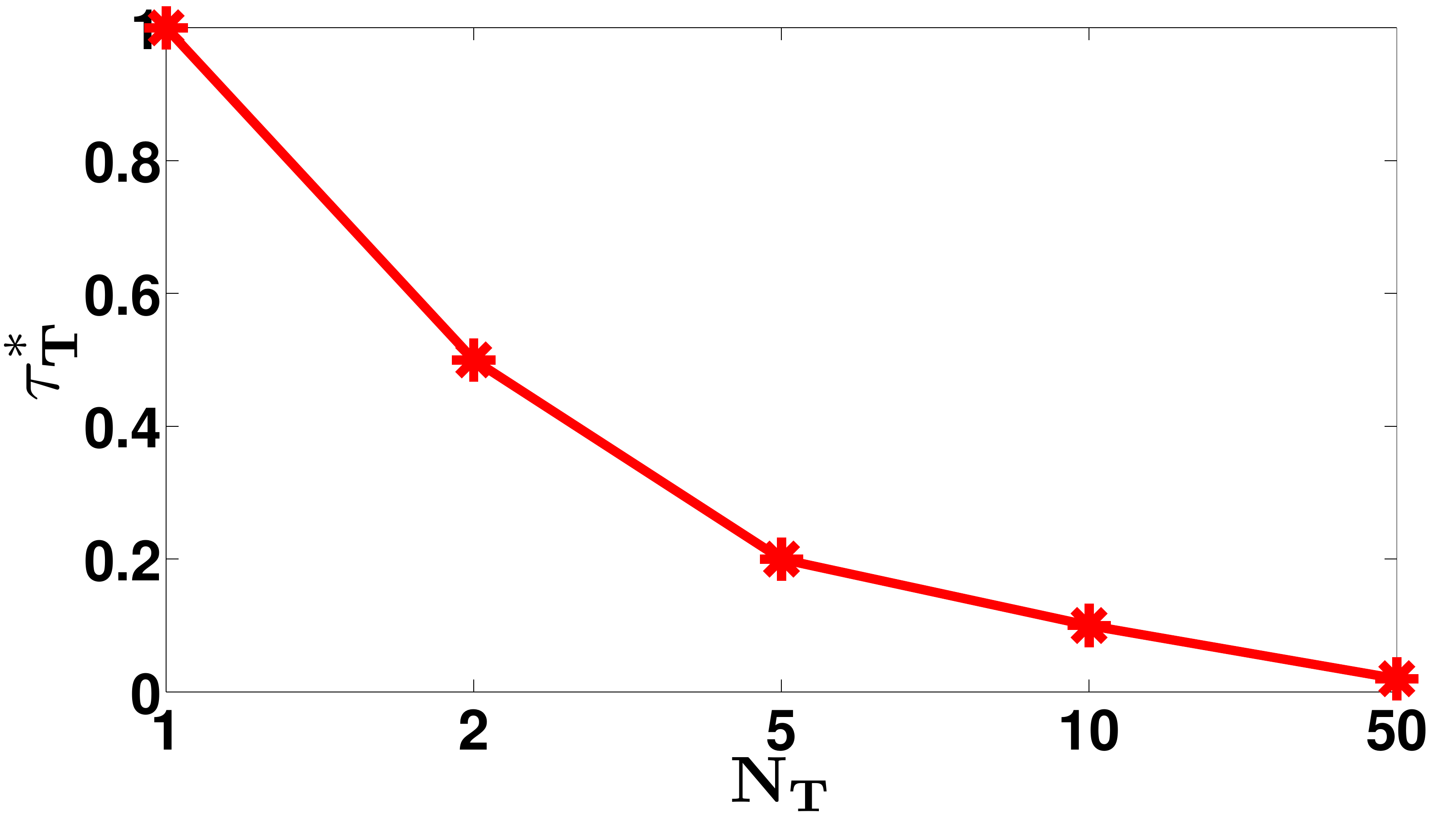}\label{fig:wifi_msne}}
  \enspace
  \subfloat[]{\includegraphics[width=0.49\columnwidth]{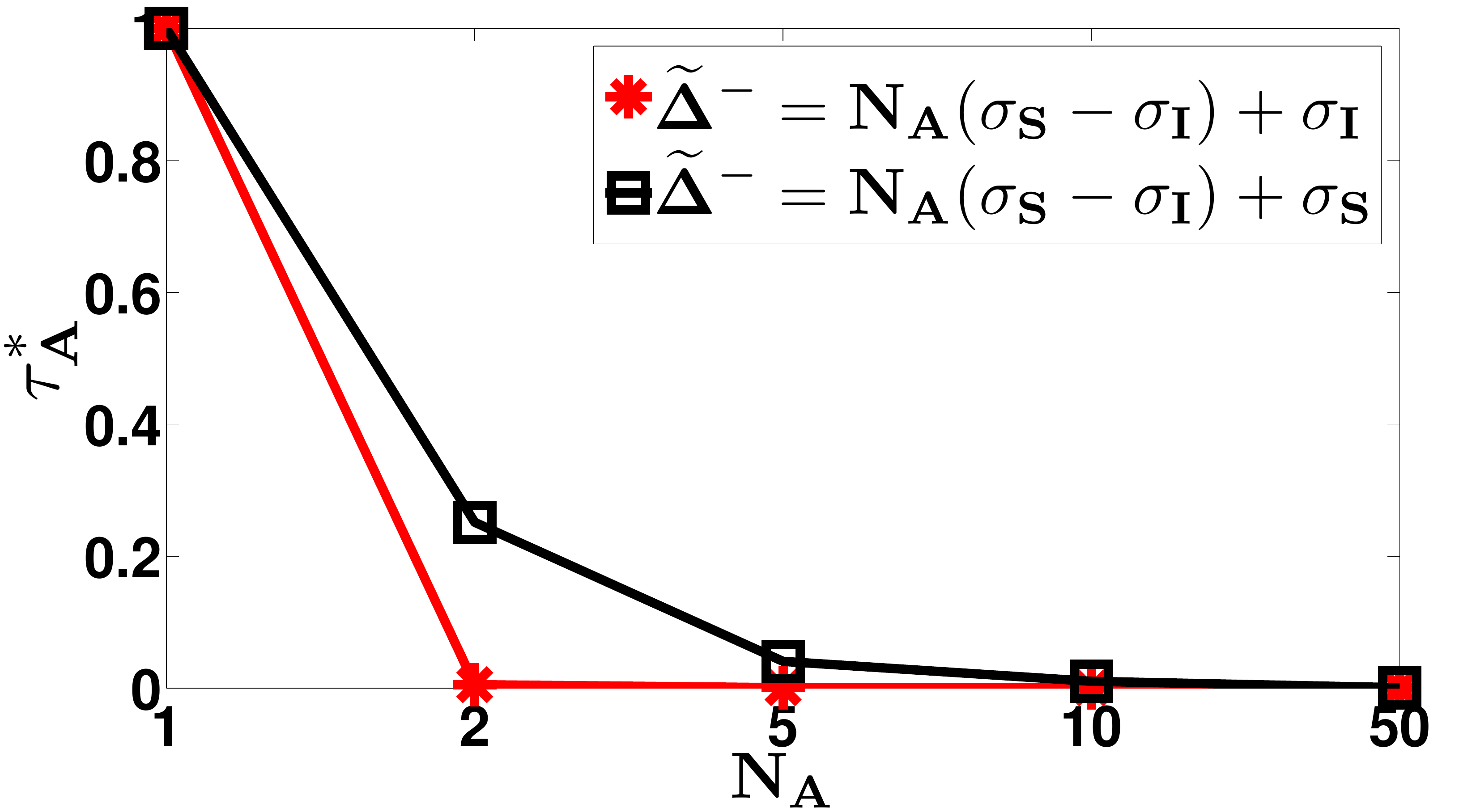}\label{fig:dsrc_msne}}
\caption{\small Access probability for (a) a $\TO$, and (b) an $\AO$ with $\AvginitAoIT{}{} = \ND(\lsucc-\lidle)+\lidle$ and  $\AvginitAoIT{}{} = \ND(\lsucc-\lidle)+\lsucc$, for different selections of $\ND$ and $\NW$.}
\label{fig:MSNEvsNodes}
\end{figure}

As stated in~\cite{zuhan}, every finite strategic-form game has a mixed strategy Nash equilibrium (MSNE). Therefore, we allow players to randomize between pure strategies and find the mixed strategy Nash equilibrium. For a strategic game $G$ defined in Section~\ref{sec:one-shot}, a mixed-strategy profile $\phi^{*} = (\phi_{\mathrm{A}}^{*},\phi_{\mathrm{T}}^{*})$ is a Nash equilibrium~\cite{zuhan}, if $\phi_{k}^{*}$ is the best response of player $k$ to his opponents' mixed strategy $\phi_{-k}^{*}\in \boldsymbol{\Phi}_{-k}$, for all $k\in\mathcal{N}$. We have
\begin{align*}
u_{k}(\phi_{k}^{*},\phi_{-k}^{*}) \geq u_{k}(\phi_{k},\phi_{-k}^{*}), \quad \forall \phi_{k} \in \boldsymbol{\Phi}_k,
\end{align*}
where 
$\phi^{*} \in \boldsymbol{\Phi} = \prod_{k=1}^{|\mathcal{N}|}\boldsymbol{\Phi}_{k}$ is the profile of mixed strategy. Since the probability distributions $\phi_{\mathrm{A}}$ and $\phi_{\mathrm{T}}$ are parameterized by $\tauD$ and $\tauW$, respectively, we find $\vect{\tau} = [\tauD^{*},\tauW^{*}]$ in order to compute the mixed strategy Nash equilibrium.
\begin{proposition}
The parameter $\vect{\tau} = [\tauD^{*},\tauW^{*}]$ required to compute the mixed strategy Nash equilibrium $\phi^{*} = (\phi_{\mathrm{A}}^{*},\phi_{\mathrm{T}}^{*})$, for the 2-player one-shot game $G$ when $\lsucc = \lcol$ is
\begin{subequations}
\begin{align}
\tauD^{*} & = 
    \begin{cases}
      \frac{\ND(\lidle-\lsucc)+\AvginitAoIT{}{}}{\ND(\lidle-\lcol+\AvginitAoIT{}{})}&\AvginitAoIT{}{} > \ND(\lsucc-\lidle), \\
      0 &\text{ otherwise }.
    \end{cases}\label{Eq:DSRC_MSNE}\\
\tauW^{*} &= \frac{1}{\NW\label{Eq:WiFi_MSNE}
}.
\end{align}
\end{subequations}
\textbf{Proof:} The proof is given in Appendix~\ref{sec:appendix_1}. 
\end{proposition}

As seen in~(\ref{Eq:DSRC_MSNE}) and (\ref{Eq:WiFi_MSNE}), the access probabilities $\tauD^{*}$ and $\tauW^{*}$ required to compute the equilibrium strategies of the $\AO$ and the $\TO$, respectively, have the following unique properties: (i) both $\tauD^{*}$ and $\tauW^{*}$ are independent of the number of nodes and access probability of the other network and (ii) $\tauD^{*}$ in any slot is a function of average age observed at the beginning of the slot i.e. $\AvginitAoIT{}{}$. As a result, the equilibrium strategy of each network is also its dominant strategy and the equilibrium strategy of the $\AO$ in any slot is a function of $\AvginitAoIT{}{}$.

Figure~\ref{fig:MSNEvsNodes} shows the $\tauW^{*}$ and $\tauD^{*}$ for the $\TO$ and the $\AO$, respectively, corresponding to different selection of nodes in the network. We show $\tauD^{*}$ for $\AvginitAoIT{}{} = \ND(\lsucc-\lidle)+\lidle$ and  $\AvginitAoIT{}{} = \ND(\lsucc-\lidle)+\lsucc$. This choice of $\AvginitAoIT{}{}$ gives $\tauD^{*}>0$ (see~(\ref{Eq:DSRC_MSNE})). Figure~\ref{fig:dsrc_msne} shows that the access probability of the $\AO$ increases from $0.0050$ to $0.2512$ with increase in $\AvginitAoIT{}{}$ from $\ND(\lsucc-\lidle)+\lidle$ to $\ND(\lsucc-\lidle)+\lsucc$ for $\ND = 2$. Also, the access probability of the $\AO$ decreases from $1$ to $0.0004$ as number of nodes in the $\AO$ increases from $1$ to $50$ for $\AvginitAoIT{}{} = \ND(\lsucc-\lidle)+\lsucc$.
\begin{figure}[t]
  \centering
  \subfloat[$\TO$ stage payoff]{\includegraphics[width=0.49\columnwidth]{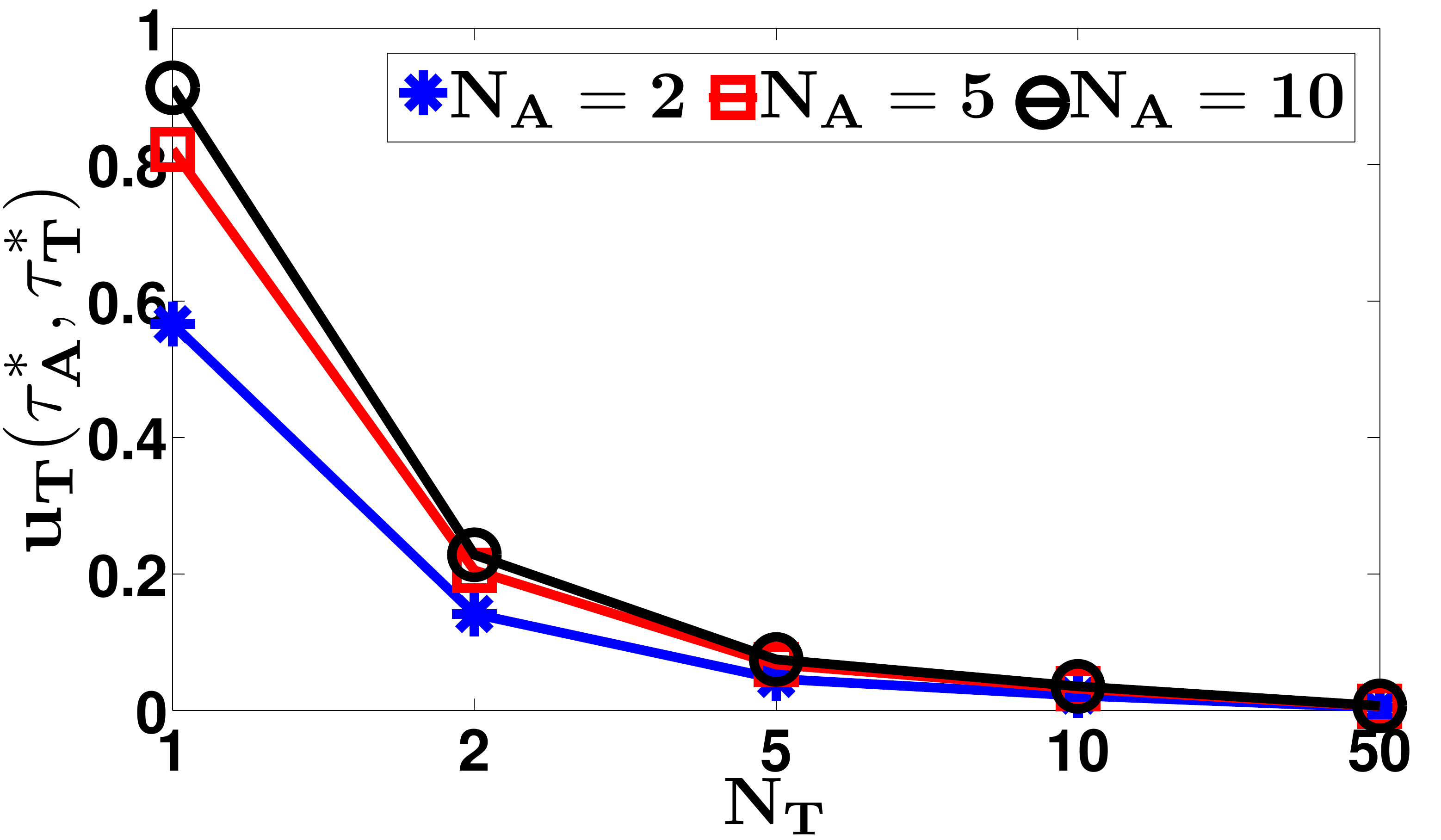}\label{fig:uw_msne}}
  \subfloat[$\AO$ stage payoff]{\includegraphics[width=0.5\columnwidth]{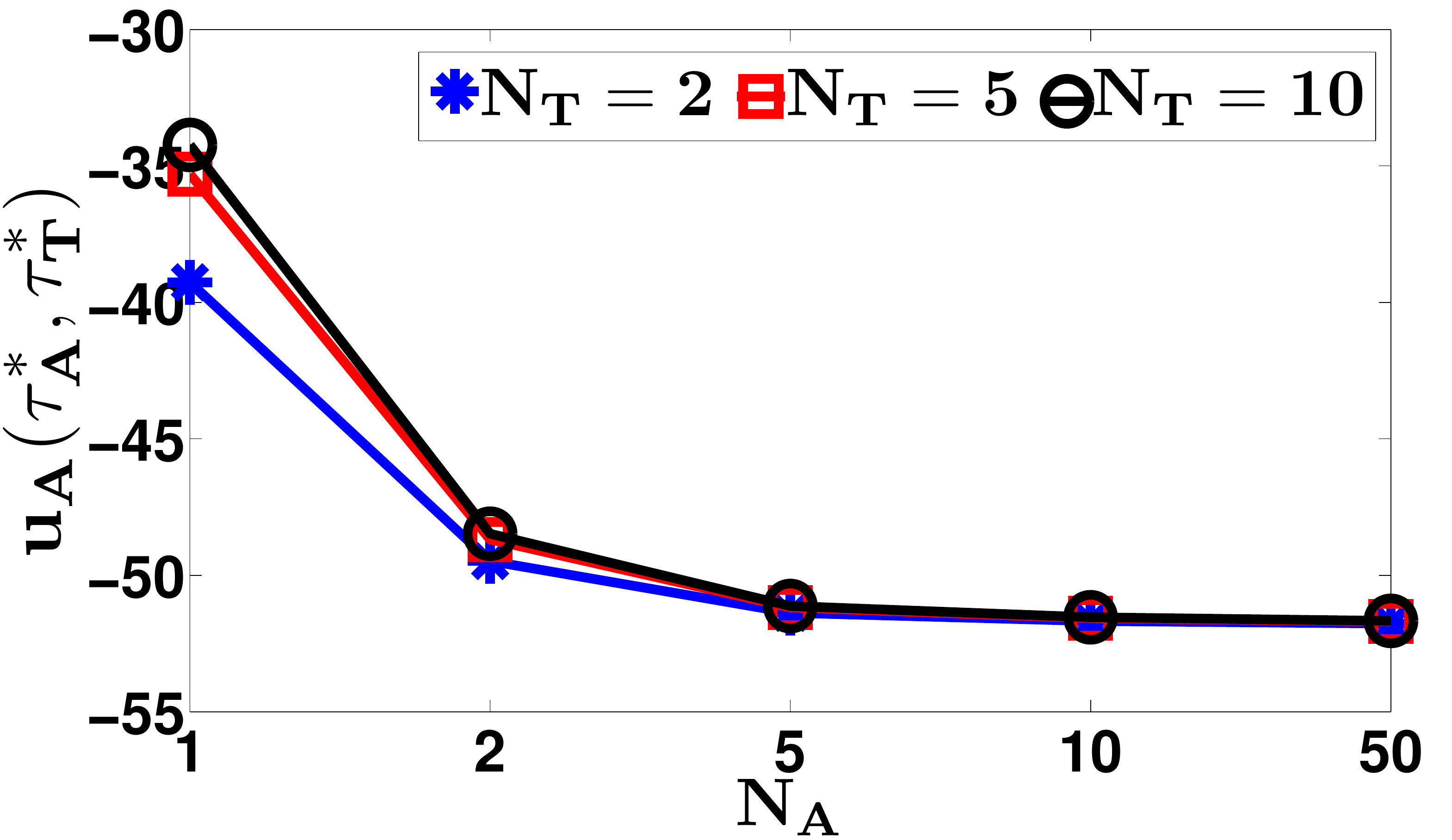}\label{fig:ud_msne}}
\caption{\small Stage payoff of a $\TO$ and an $\AO$ for different selections of $\NW$ and $\ND$. The $\AO$ stage utilities correspond to $\AvginitAoIT{}{} = \ND(\lsucc-\lidle)+\lsucc$.}
\label{fig:PayoffvsNodes}
\end{figure}

\begin{figure*}[t]
\subfloat[Throughput]{\includegraphics[width = 0.31\textwidth]{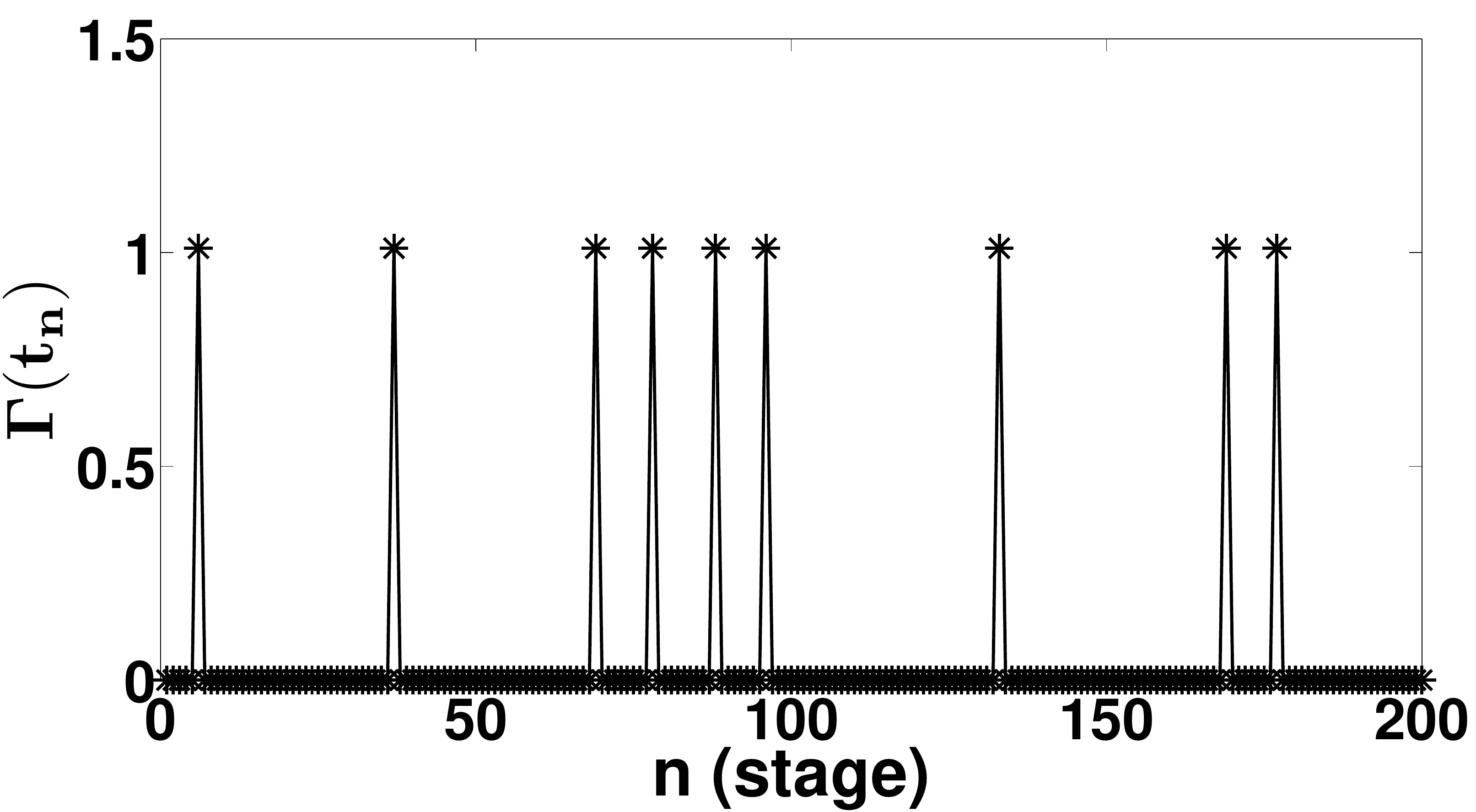}\label{fig:Ex_WiFi}}
\quad
\subfloat[Age of update]{\includegraphics[width = 0.31\textwidth]{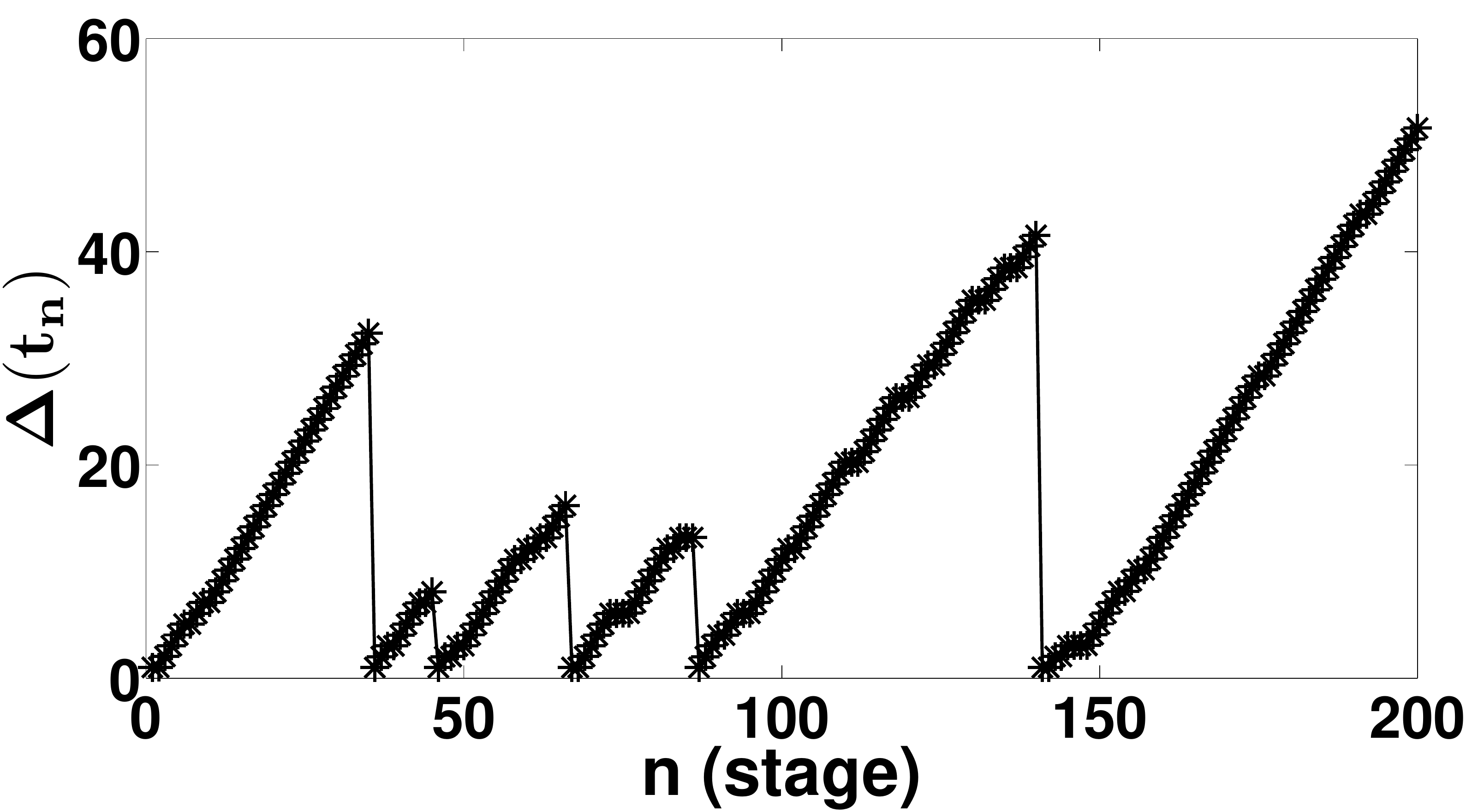}\label{fig:Ex_DSRC}}
\quad
\subfloat[Access probability]{\includegraphics[width = 0.32\textwidth]{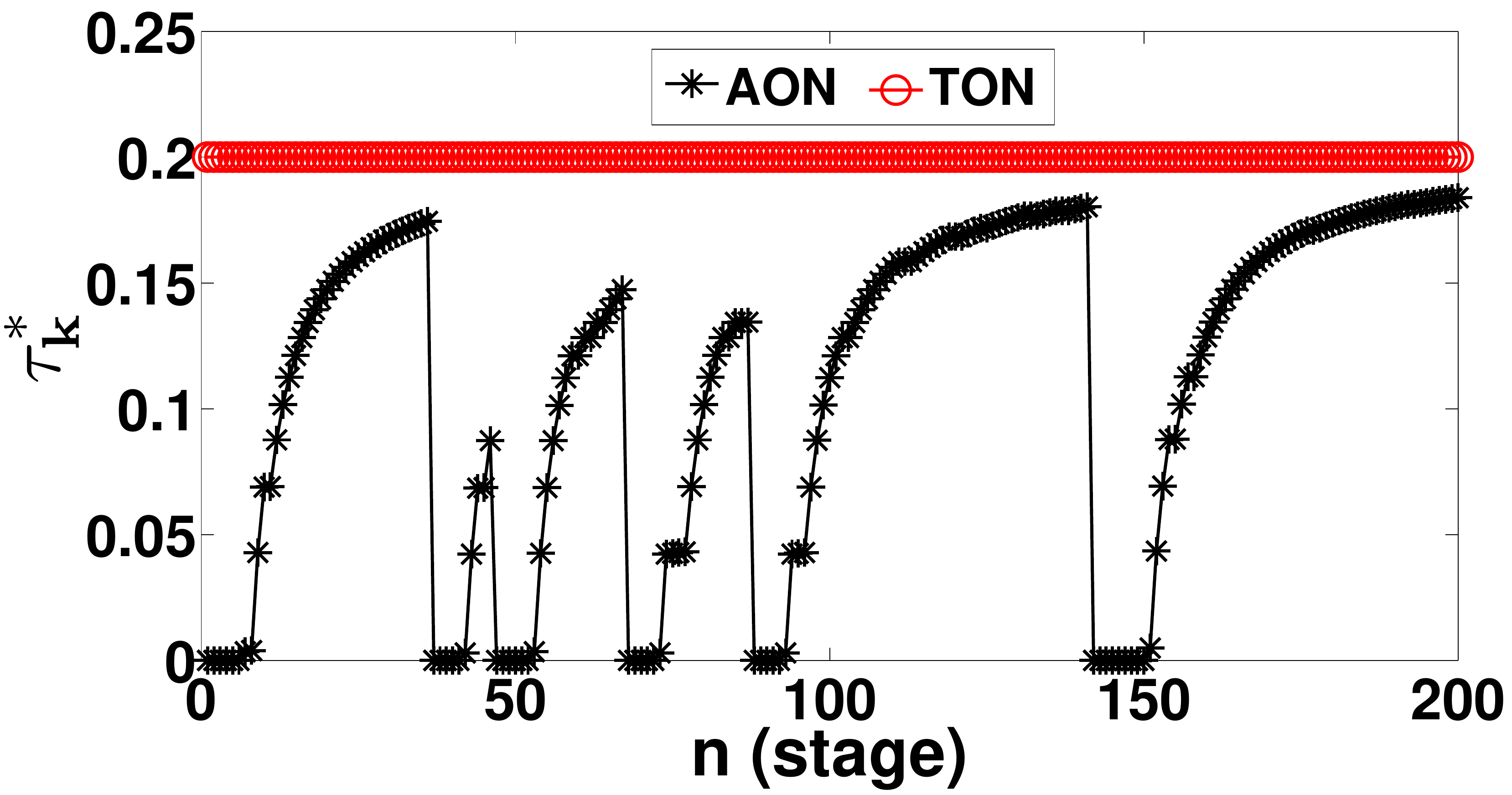}\label{fig:Ex_NE}}
\caption{\small Illustration of per stage (a) throughput of a $\TO$ (b) age of an $\AO$ and (c) access probability of an $\AO$ and a $\TO$ wrt stage obtained from an independent run. The results correspond to $\ND = 5, \NW = 5, \lsucc = \lcol = 1+\beta, \lidle = \beta$ and $\beta = 0.01$.}
\label{fig:ExRun}
\end{figure*}

A distinct feature of the stage game is the effect of \textit{self-contention} and \textit{competition} on the network utilities which we had also observed in our earlier work~\cite{SGAoI2018}. We define self-contention as the impact of nodes within one's own network and competition as the impact of nodes in the other network. Figure~\ref{fig:PayoffvsNodes} shows the affect of self-contention and competition on the network utilities, when networks play their respective equilibrium strategies. As shown in Figure~\ref{fig:uw_msne}, as the number of nodes in the $\AO$ increase, the payoff of the $\TO$ increases. Intuitively, since increase in the number of $\AO$ nodes results in increase in competition, the payoff of the $\TO$ should decrease. However, the payoff of the $\TO$ increases. For example, for $\NW = 2$, as shown in Figure~\ref{fig:uw_msne}, the payoff of the $\TO$ increases from $0.1416$ to $0.2281$ as $\ND$ increases from $2$ to $10$. This increase is due to increase in self-contention within the $\AO$ which forces the network to be conservative and hence benefits the $\TO$. Similarly as shown in Figure~\ref{fig:ud_msne}, as the number of $\TO$ nodes increases the $\AO$ payoff improves. 
\subsection{Repeated game}
\label{sec:repeated_game}

As the networks coexist over a long period of time, the one-shot game defined in Section~\ref{sec:one-shot} is played in every stage (slot) $n\in\{1,2,\dots\}$. We consider an infinitely repeated game, defined as $G^{\infty}$ with perfect monitoring~\cite{zuhan} i.e. at the end of each stage, all players observe the action profile chosen by every other player\footnote{Assumptions such as imperfect and private monitoring are more realistic. However, for ease of exposition, we assume perfect monitoring and propose to study coexistence under other monitoring assumptions in the future.}. In addition to the action profiles, players also observe $\AvginitAoIT{n}{}$ i.e. the average age of the $\AO$ at the end of stage $(n-1)$. We refer to $\AvginitAoIT{n}{}$ as the state variable. \SG{A feasible strategy of the repeated game, in general, would depend on the history of play and the state variable. However, here we restrict ourselves to studying the subgame perfect equilibria that involves the simplest kind of strategies i.e. players play the MSNE in each stage. In a repeated game with no state variable such a strategy would perhaps be uninteresting. However, our game is a repeated game where the $\AO$ equilibrium strategy as shown in~(\ref{Eq:DSRC_MSNE}) in any stage is a function of the state variable. The dependence of the $\AO$ equilibrium strategy on the state variable intertwines the utilities of the networks, even though the equilibrium strategy of each network is independent of the other network and \SG{allows} us to explore interesting aspects of the game.}

Figure~\ref{fig:ExRun} shows the payoffs (see~(\ref{Eq:thrpayoff})-(\ref{Eq:agepayoff})) and the access probabilities of the $\TO$ and the $\AO$ for the repeated game $G^{\infty}$. The results correspond to a $\AO$-$\TO$ coexistence with $\ND = \NW = 5$. Figure~\ref{fig:Ex_DSRC} and~\ref{fig:Ex_NE}, illustrating the evolution of age and access probability of the $\AO$, are interlinked. As shown in~(\ref{Eq:DSRC_MSNE}), $\tauD^{*}$, is a function of age observed in the beginning of a stage. The threshold value i.e. $\ND(\lsucc - \lidle)$, for $\ND = \NW = 5$, $\lsucc = 1+\beta$, $\lidle = \beta$ and $\beta = 0.01$ is $5$. As a result, nodes in the $\AO$ access the medium with $\tauD^* > 0$ in any stage $n$ only if the average age in the $(n-1)^{th}$ stage exceeds a threshold value i.e. $\AvginitAoIT{n}{}> 5$, otherwise $\tauD^{*} = 0$. For instance, in Figure~\ref{fig:Ex_NE}, $\tauD^{*} = 0$ for $n \in [37, 41]$ since $\AvginitAoIT{n}{}< 5$, however, for $n =42$, $\tauD^{*} = 0.0030$ as $\AvginitAoIT{42}{} =  6.0700$ exceeds the threshold value.

Player $k$'s average discounted payoff for the game $G^{\infty}$, where $k \in \mathcal{N}$ is
\begin{align}
U_{k}= E_{\phi}\left\{(1-\alpha)\sum\limits_{n=1}^{\infty}\alpha^{n-1}u_{k}(\phi)\right\}.
\label{Eq:AvgDisPayoff}
\end{align}
where, the expectation is taken with respect to the strategy profile $\phi$, $u_{k}(\phi)$ is player $k$'s payoff in stage $n$ and $0<\alpha<1$ is the discount factor. Note that a discount factor $\alpha$ closer to $1$ means that the player values not only the stage payoff but also the payoff in the future i.e. the player is far-sighted, whereas $\alpha$ closer to $0$ means that the player is myopic and values only the current payoff. By substituting~(\ref{Eq:thrpayoff}) and~(\ref{Eq:agepayoff}) in~(\ref{Eq:AvgDisPayoff}), we can obtain the average discounted payoffs $U_{W}$ and $U_{D}$ of the $\TO$ and the $\AO$, respectively.
\section{Results}
\label{sec:results}
In this section, we first discuss the simulation setup and later the results. For what follows, we set $\lidle = \beta$, $0 < \beta < 1$, and $\lsucc = \lcol = (1+\beta)$. In practice, the idle slot is much smaller than a collision or a successful transmission slot, that is, $\beta << 1$. We select $\beta = 0.01$ for the simulation results discussed ahead. We make different selections of $\ND$ and $\NW$ to illustrate the impact of self-contention and competition. Specifically, we simulate for $\ND \in \{1, 2, 5, 10, 50\}$ and $\NW \in \{1, 2, 5, 10, 50\}$. We consider $\alpha \in [0.01,0.99]$ for computing the discounted payoffs. Different selections of $\alpha$ allow us to study the behavior of myopic and far-sighted players. We use Monte Carlo simulations to compute the average discounted payoff of the $\AO$ and the $\TO$. We compute the average over $100,000$ independent runs each comprising of $1000$ stages. For each run, we consider the initial age $\AvginitAoIT{1}{} = \lsucc = (1+\beta)$. Also, we fix the rate of transmission $r$ for each node in the WiFi network to $1$ bit/sec.

\begin{figure}[t]
\subfloat[$\TO$ Expected Payoff~(\ref{Eq:AvgDisPayoff})]{\includegraphics[width = 0.49\columnwidth]{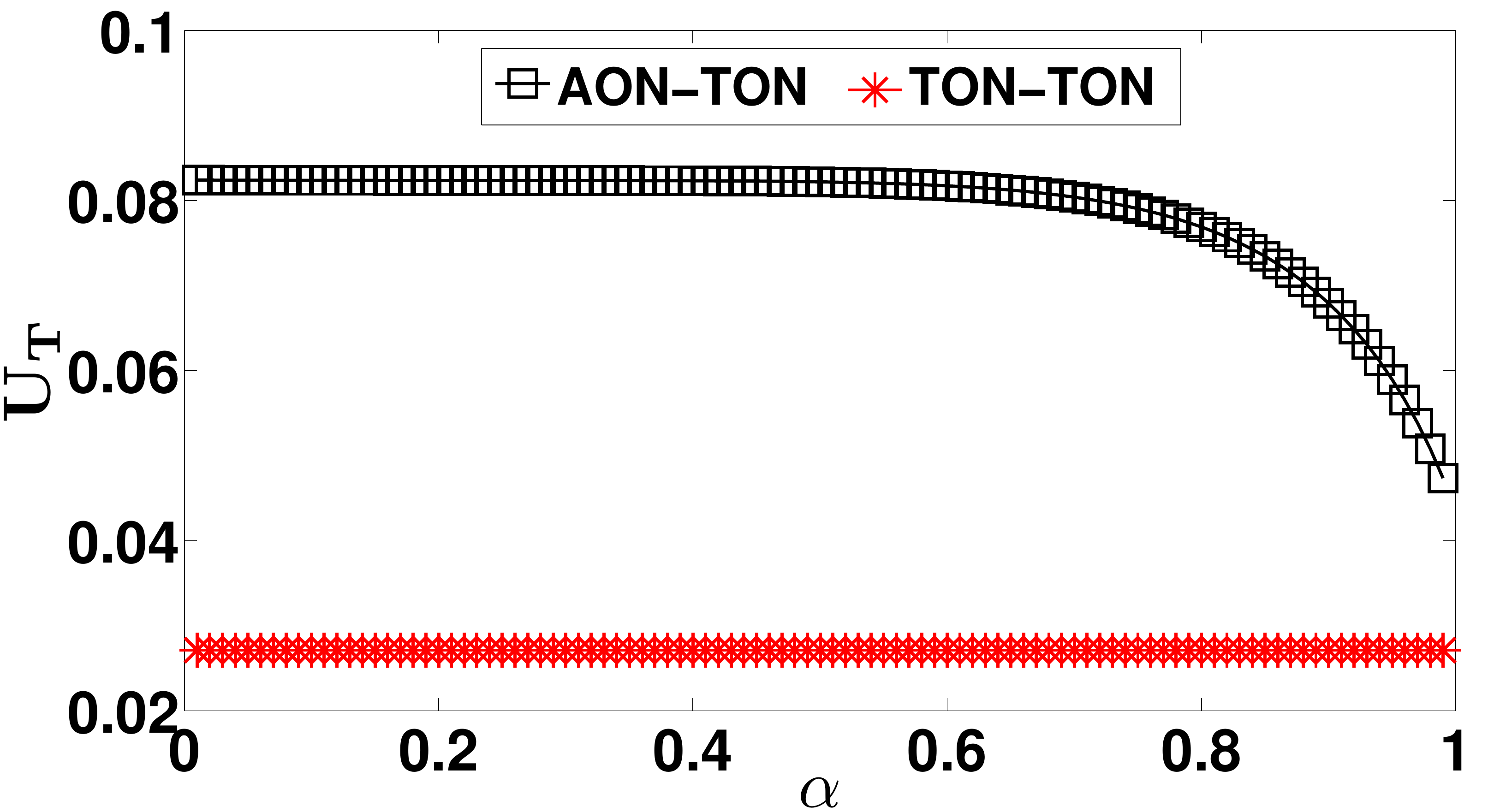}\label{fig:DDvsDWvsWW_WiFiPayoff}}
\enspace
\subfloat[$\AO$ Expected Payoff~(\ref{Eq:AvgDisPayoff})]{\includegraphics[width = 0.48\columnwidth]{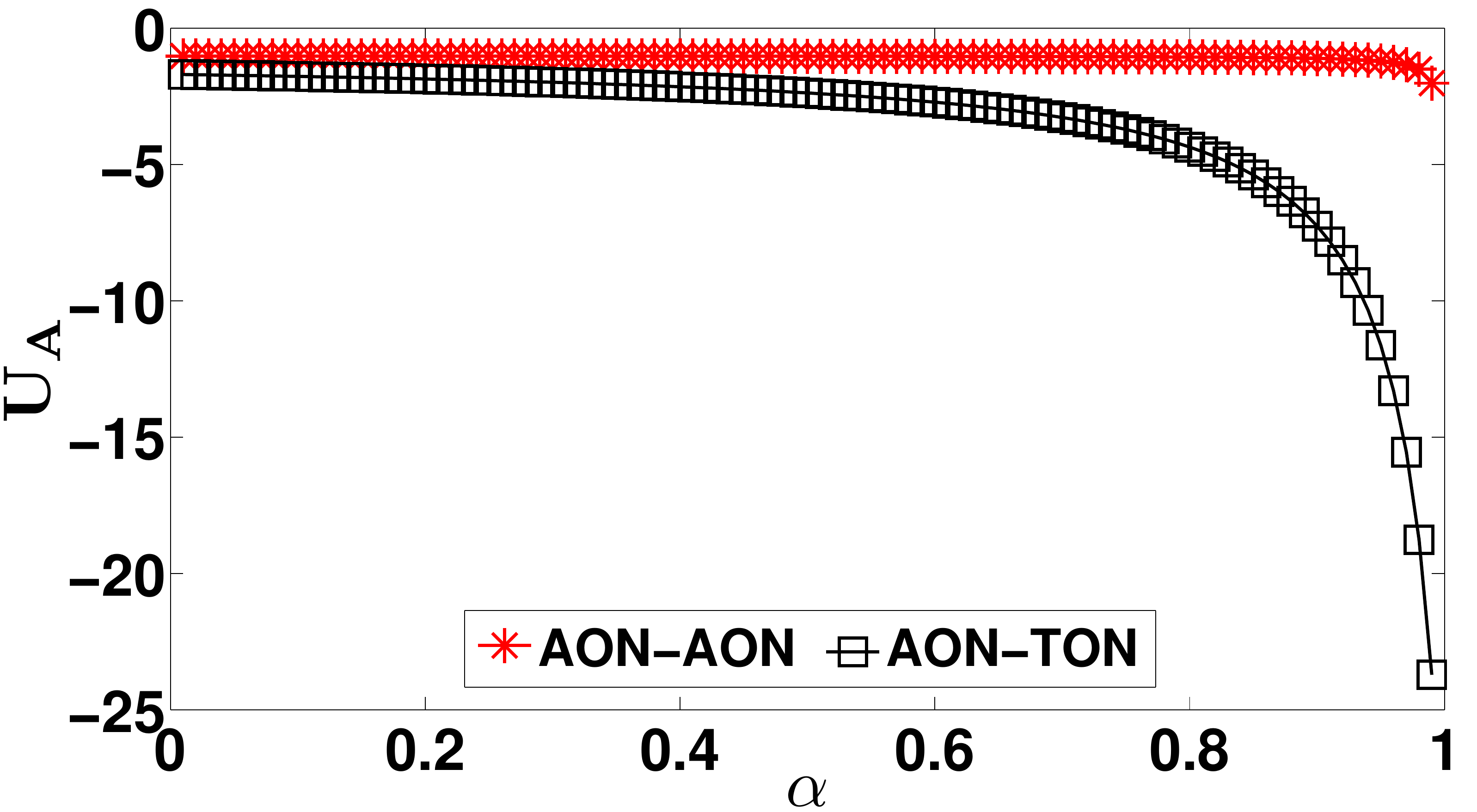}\label{fig:DDvsDWvsWW_DSRCPayoff}}
\caption{\small Discounted payoff of the $\TO$ and the $\AO$ to illustrate the impact of coexistence. Networks under study comprises of $5$ nodes each.}
\label{fig:DDvsDWvsWW}
\end{figure}

To understand the impact of coexistence on the $\AO$ and the $\TO$, we consider three coexistence scenarios: (i) an $\AO$ coexists with a $\TO$, (ii) an $\AO$ coexists with another $\AO$ and (iii) a $\TO$ coexists with another $\TO$. Similar to $\AO$-$\TO$ coexistence, networks in $\AO$-$\AO$ and $\TO$-$\TO$ coexistence randomize between pure strategies\footnote{The parameter $\tau_{k}^{*}$, where $k\in \{\mathrm{A}_{\mathrm{I}},\mathrm{A}_{\mathrm{II}}\}$, required to compute the mixed strategy Nash equilibrium $\phi^{*} = (\phi_{\mathrm{A}_{\mathrm{I}}}^{*},\phi_{\mathrm{A}_{\mathrm{II}}}^{*})$, for $\AO$-$\AO$ coexistence with $N_{\mathrm{A}_{\mathrm{I}}}$ and $N_{\mathrm{A}_{\mathrm{II}}}$ nodes in $\AO$ $\mathrm{I}$ and $\mathrm{II}$, respectively, is
\begin{align*}
\tau_{k}^{*} & = 
    \begin{cases}
      \frac{N_{k}(\lidle-\lsucc)+\AvginitAoIT{}{}}{N_{k}(\lidle-\lcol+\AvginitAoIT{}{})}&\text{if } \AvginitAoIT{}{} > N_{k}(\lsucc-\lidle), \\
      0 &\text{ otherwise }.
    \end{cases}
\end{align*}
Similarly, the parameter $\tau_{k}^{*}$, where $k\in \{\mathrm{T}_{\mathrm{I}},\mathrm{T}_{\mathrm{II}}\}$, required to compute the mixed strategy Nash equilibrium $\phi^{*} = (\phi_{\mathrm{T}_{\mathrm{I}}}^{*},\phi_{\mathrm{T}_{\mathrm{II}}}^{*})$, for $\TO$-$\TO$ coexistence with $N_{\mathrm{T}_{\mathrm{I}}}$ and $N_{\mathrm{T}_{\mathrm{II}}}$ nodes in $\TO$ $\mathrm{I}$ and $\mathrm{II}$, respectively, is $\tau_{k}^{*} = 1/N_{k}$.}. Figure~\ref{fig:DDvsDWvsWW} shows the discounted payoff with respect to the discount factor $\alpha$ for the above coexistence scenario. We now discuss the impact of $\AO$ on $\TO$ payoff and $\TO$ on $\AO$ payoff in detail.

\textit{Impact of $\AO$ on $\TO$ payoff:} As shown in Figure~\ref{fig:DDvsDWvsWW_WiFiPayoff}, a $\TO$ sees significant improvement in payoff when the coexisting network is an $\AO$. This improvement in $\TO$ payoff is due to the dynamic nature of the $\AO$ equilibrium strategy~(\ref{Eq:DSRC_MSNE}). The empirical frequency of occurence of $\tauD^{*} = 0$ when an $\AO$ coexists with a $\TO$ and each network has $5$ nodes is $0.13$. This means that a $\TO$ gets an additional $13\%$ stages to transmit without any competition from the $\AO$ in $\AO$-$\TO$ coexistence as compared to $\TO$-$\TO$ coexistence. Consequently, the empirical frequency of $\TO$ seeing a successful transmission increases from $0.027$ as seen in $\TO$-$\TO$ coexistence to $0.043$ in $\AO$-$\TO$ coexistence. Also, the empirical frequency of failed transmissions (collision) decreases from $0.624$ as seen in $\TO$-$\TO$ coexistence to $0.017$ in $\AO$-$\TO$ coexistence. Table~\ref{tab:DDvsDWvsWW} shows the empirical frequency of successful transmission, collision and $\tauD^{*} = 0$ for different scenarios under study.

\begin{figure}[t]
\subfloat[]{\includegraphics[width = 0.48\columnwidth]{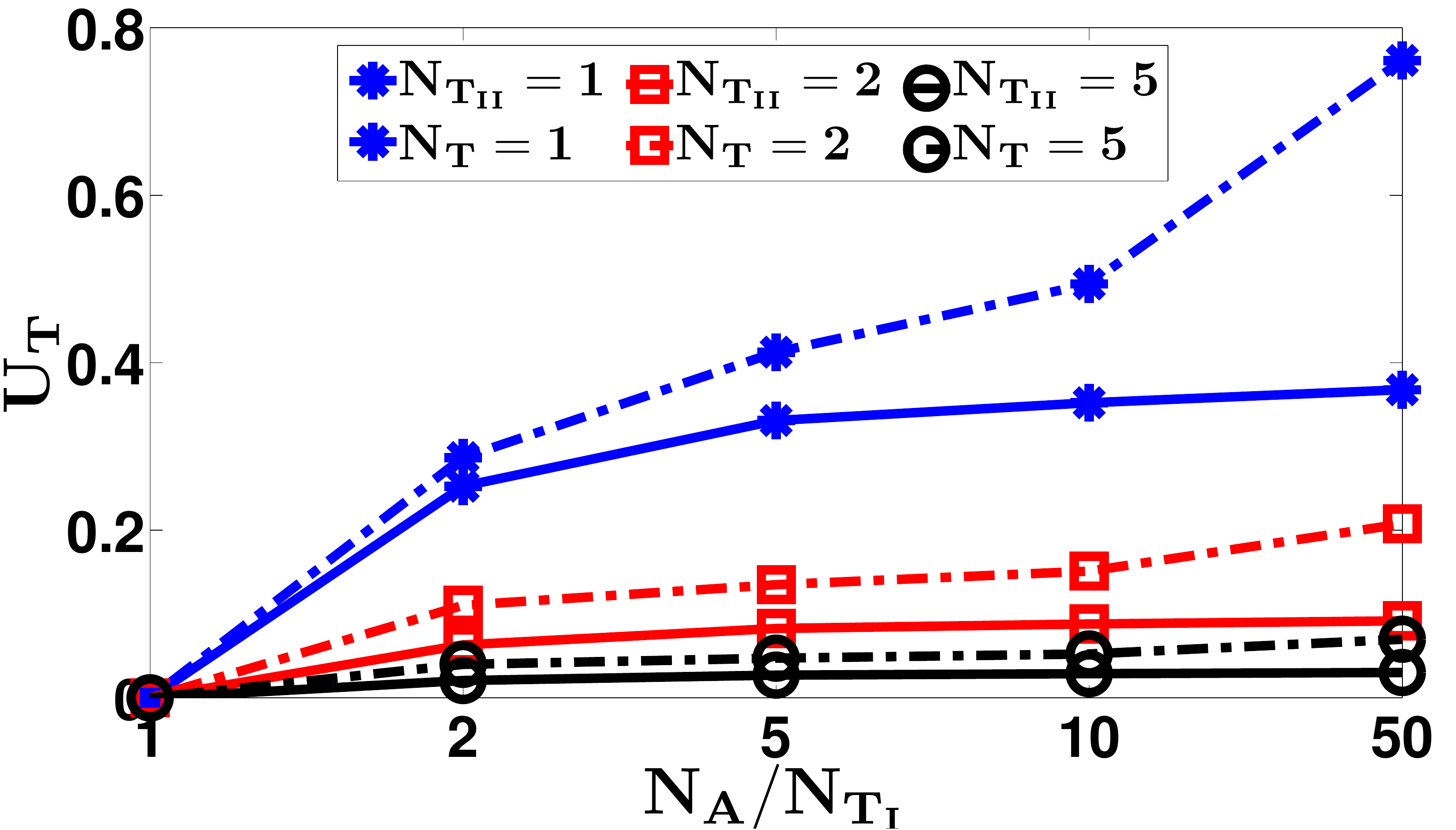}\label{fig:UW}}
\enspace
\subfloat[]{\includegraphics[width = 0.49\columnwidth]{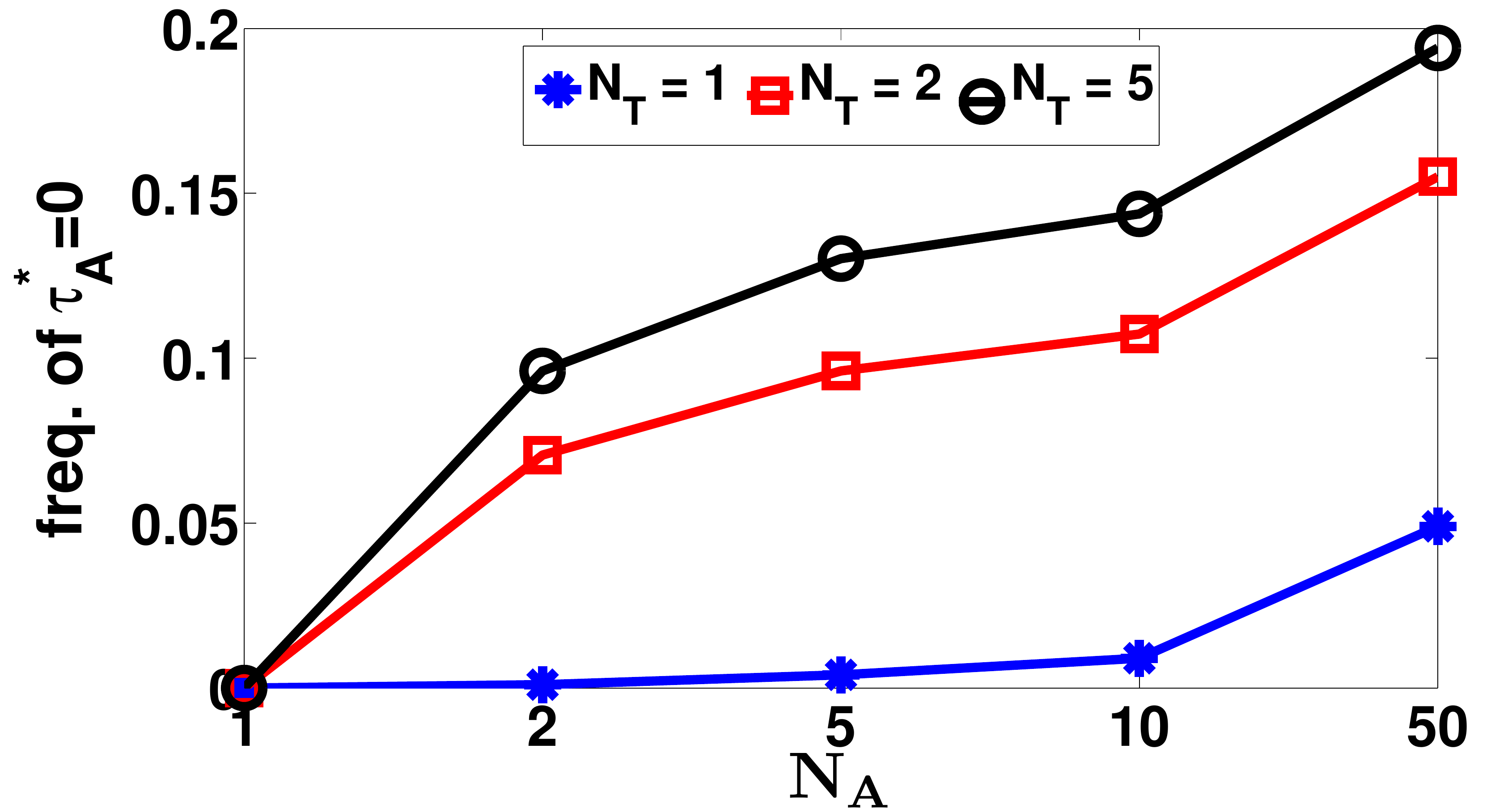}\label{fig:NT}}
\caption{\small Variation in (a) discounted payoff of $\TO$ in $\TO$-$\TO$ (solid line) and $\AO$-$\TO$ (dashed line) coexistence scenario and (b) frequency of $\tauD^{*} = 0$, with respect to increasing number of $\AO$ nodes. Computations are done for $\beta = 0.01$ and $\alpha = 0.99$.}
\label{fig:probs}
\end{figure}

Figure~\ref{fig:UW} shows the discounted payoff of a $\TO$ for both $\AO$-$\TO$ and $\TO$-$\TO$ coexistence. As shown in Figure~\ref{fig:UW}, the benefits for a $\TO$ in $\AO$-$\TO$ coexistence further increases, as the size of the $\AO$ network increases. This is due to the increase in \textit{self-contention} within the $\AO$, which forces it to be conservative. As shown in~(\ref{Eq:DSRC_MSNE}), the age threshold $\ND(\lsucc-\lidle)$ increases with increase in $\ND$. As a result, the frequency of occurence of $\tauD^{*} = 0$ increases. We illustrate the increase in the frequency of occurence of $\tauD^{*} = 0$ with increasing number of $\AO$ nodes in Figure~\ref{fig:NT}. This increase in the frequency of occurence of $\tauD^{*} = 0$ works in favour of a $\TO$.
\begin{table}
\ra{1}
\centering
\caption{\small{Empirical frequency of successful transmission, collision and occurence of $\tauD^{*} = 0$ for different coexistence scenarios computed over $100,000$ independent runs with $1000$ stages each. Networks under study comprises of $5$ nodes each and the calculations are done for $\beta = 0.01$.}}
\resizebox{\columnwidth}{!}{%
{\large
\begin{tabular}{lcccc}\toprule
\parbox[t]{2.5cm}{\textbf{Coexistence scenario (I-II)}}& \parbox[t]{2.5cm}{\textbf{Frequency of successful transmission in Network I}} & \parbox[t]{2.5cm}{\textbf{Frequency of successful transmission in Network II}} & \parbox[t]{2.5cm}{\textbf{Frequency of collision}} & \parbox[t]{2.5cm}{\textbf{Frequency of $\mathbf{\tauD^{*} = 0}$ \\(I, II)}}\\
\midrule
$\AO$-$\AO$ & $0.004$ & $0.004$ & $0.002$ & $0.877,0.877$\\
$\AO$-$\TO$ & $0.021$ & $0.043$ & $0.017$ & $0.13, \text{NA}$\\
$\TO$-$\TO$ & $0.027$ & $0.027$ & $0.624$ & $\text{NA, NA}$\\
\bottomrule
\end{tabular}}}
\label{tab:DDvsDWvsWW}
\end{table}

\textit{Impact of $\TO$ on $\AO$ payoff:} As shown in Figure~\ref{fig:DDvsDWvsWW_DSRCPayoff}, an $\AO$ sees a larger age when it coexists with a $\TO$ as compared to when it coexists with another $\AO$. The frequency of occurence of $\tauD^{*} = 0$ as shown in Table~\ref{tab:DDvsDWvsWW} is $0.877$ and $0.13$ for $\AO$-$\AO$ and $\AO$-$\TO$ coexistence, respectively. While the frequency of occurence of $\tauD^{*} = 0$ in $\AO$-$\AO$ coexistence is higher than in $\AO$-$\TO$ coexistence, the age in the former coexistence scenario is still smaller than that in the latter. This is due to the increase in contention from the $\TO$, which has a static equilibrium strategy $\tauW^{*} = 1/N_{W}$ and which increases the probability of collision from $0.002$ in $\AO$-$\AO$ coexistence to $0.017$ in $\AO$-$\TO$ coexistence as shown in Table~\ref{tab:DDvsDWvsWW}. Therefore, spectrum sharing with a $\TO$ is detrimental for an $\AO$ as compared to sharing with another $\AO$.
\section{Conclusion}
\label{sec:conclusion}
We formulated a repeated game to study the coexistence problem between age and throughput optimizing networks. We characterized the mixed strategy Nash equilibrium of the stage game and studied the evolution of the equilibrium strategies over time and the resulting average discounted payoffs of the networks. We showed that unlike $\TO$-$\TO$ coexistence where nodes in both the networks access the medium aggressively to maximize their respective throughputs, in $\AO$-$\AO$ coexistence, the requirement of timely updates of the $\AO$ makes it conservative and occasionally refrains its nodes from accessing the medium. This works in favour of the $\TO$, therefore, making spectrum sharing with an $\AO$ beneficial for a $\TO$ in comparison to when a $\TO$ shares the medium with another $\TO$. In addition, we showed that spectrum sharing with a $\TO$ is detrimental to an $\AO$ in comparison to sharing with another $\AO$.
\begin{spacing}{0.95}
\bibliographystyle{IEEEtran}
\bibliography{references}

\begin{thebibliography}{10}
\providecommand{\url}[1]{#1}
\csname url@samestyle\endcsname
\providecommand{\newblock}{\relax}
\providecommand{\bibinfo}[2]{#2}
\providecommand{\BIBentrySTDinterwordspacing}{\spaceskip=0pt\relax}
\providecommand{\BIBentryALTinterwordstretchfactor}{4}
\providecommand{\BIBentryALTinterwordspacing}{\spaceskip=\fontdimen2\font plus
\BIBentryALTinterwordstretchfactor\fontdimen3\font minus
  \fontdimen4\font\relax}
\providecommand{\BIBforeignlanguage}[2]{{%
\expandafter\ifx\csname l@#1\endcsname\relax
\typeout{** WARNING: IEEEtran.bst: No hyphenation pattern has been}%
\typeout{** loaded for the language `#1'. Using the pattern for}%
\typeout{** the default language instead.}%
\else
\language=\csname l@#1\endcsname
\fi
#2}}
\providecommand{\BIBdecl}{\relax}
\BIBdecl

\bibitem{fanet}
I.~Bekmezci, O.~K. Sahingoz, and {\c{S}}.~Temel, ``Flying ad-hoc networks
  (fanets): A survey,'' \emph{Ad Hoc Networks}, vol.~11, no.~3, pp. 1254--1270,
  2013.

\bibitem{vanet}
H.~Hartenstein and L.~Laberteaux, ``A tutorial survey on vehicular ad hoc
  networks,'' \emph{IEEE Communications magazine}, vol.~46, no.~6, pp.
  164--171, 2008.

\bibitem{liu2017}
J.~Liu, G.~Naik, and J.-M.~J. Park, ``Coexistence of dsrc and wi-fi: Impact on
  the performance of vehicular safety applications,'' in \emph{Communications
  (ICC), 2017 IEEE International Conference on}.\hskip 1em plus 0.5em minus
  0.4em\relax IEEE, 2017, pp. 1--6.

\bibitem{kaul2011minimizing}
S.~Kaul, M.~Gruteser, V.~Rai, and J.~Kenney, ``Minimizing age of information in
  vehicular networks,'' in \emph{Sensor, Mesh and Ad Hoc Communications and
  Networks (SECON), 2011 8th Annual IEEE Communications Society Conference
  on}.\hskip 1em plus 0.5em minus 0.4em\relax IEEE, 2011, pp. 350--358.

\bibitem{mario2005}
M.~Cagalj, S.~Ganeriwal, I.~Aad, and J.-P. Hubaux, ``On selfish behavior in
  csma/ca networks,'' in \emph{INFOCOM 2005. 24th Annual Joint Conference of
  the IEEE Computer and Communications Societies. Proceedings IEEE},
  vol.~4.\hskip 1em plus 0.5em minus 0.4em\relax IEEE, 2005, pp. 2513--2524.

\bibitem{naik2017}
G.~Naik, J.~Liu, and J.-M.~J. Park, ``Coexistence of dedicated short range
  communications (dsrc) and wi-fi: Implications to wi-fi performance,'' in
  \emph{Proc. IEEE INFOCOM}, 2017.

\bibitem{UAV_delay}
M.~Asadpour, D.~Giustiniano, K.~A. Hummel, S.~Heimlicher, and S.~Egli, ``Now or
  later?: Delaying data transfer in time-critical aerial communication,'' in
  \emph{Proceedings of the ninth ACM conference on Emerging networking
  experiments and technologies}.\hskip 1em plus 0.5em minus 0.4em\relax ACM,
  2013, pp. 127--132.

\bibitem{UAV_throughput_2}
S.~Hayat, E.~Yanmaz, and C.~Bettstetter, ``Experimental analysis of
  multipoint-to-point uav communications with ieee 802.11 n and 802.11 ac,'' in
  \emph{2015 IEEE 26th Annual International Symposium on Personal, Indoor, and
  Mobile Radio Communications (PIMRC)}.\hskip 1em plus 0.5em minus 0.4em\relax
  IEEE, 2015, pp. 1991--1996.

\bibitem{sun2017update}
Y.~Sun, E.~Uysal-Biyikoglu, R.~D. Yates, C.~E. Koksal, and N.~B. Shroff,
  ``Update or wait: How to keep your data fresh,'' \emph{IEEE Transactions on
  Information Theory}, 2017.

\bibitem{yates2015lazy}
R.~D. Yates, ``Lazy is timely: Status updates by an energy harvesting source,''
  in \emph{Information Theory (ISIT), 2015 IEEE International Symposium
  on}.\hskip 1em plus 0.5em minus 0.4em\relax IEEE, 2015, pp. 3008--3012.

\bibitem{chen2010}
L.~Chen, S.~H. Low, and J.~C. Doyle, ``Random access game and medium access
  control design,'' \emph{IEEE/ACM Transactions on Networking (TON)}, vol.~18,
  no.~4, pp. 1303--1316, 2010.

\bibitem{impact2017}
G.~D. Nguyen, S.~Kompella, C.~Kam, J.~E. Wieselthier, and A.~Ephremides,
  ``Impact of hostile interference on information freshness: A game approach,''
  in \emph{Modeling and Optimization in Mobile, Ad Hoc, and Wireless Networks
  (WiOpt), 2017 15th International Symposium on}.\hskip 1em plus 0.5em minus
  0.4em\relax IEEE, 2017, pp. 1--7.

\bibitem{YinAoI2018}
Y.~Xiao and Y.~Sun, ``A dynamic jamming game for real-time status updates,''
  \emph{arXiv preprint arXiv:1803.03616}, 2018.

\bibitem{SGAoI2018}
S.~Gopal and S.~K. Kaul, ``A game theoretic approach to dsrc and wifi
  coexistence,'' in \emph{IEEE INFOCOM 2018 - IEEE Conference on Computer
  Communications Workshops (INFOCOM WKSHPS)}, April 2018, pp. 565--570.

\bibitem{bianchi}
G.~Bianchi, ``Performance analysis of the ieee 802.11 distributed coordination
  function,'' \emph{IEEE Journal on selected areas in communications}, vol.~18,
  no.~3, pp. 535--547, 2000.

\bibitem{zuhan}
Z.~Han, D.~Niyato, W.~Saad, T.~Ba{\c{s}}ar, and A.~Hj{\o}rungnes, \emph{Game
  Theory in Wireless and Communication Networks: Theory, Models, and
  Applications}.\hskip 1em plus 0.5em minus 0.4em\relax Cambridge University
  Press, 2011.

\bibitem{SGAoI2019}
S.~Gopal, S.~K. Kaul, and R.~Chaturvedi, ``Coexistence of age and throughput
  optimizing networks: A game theoretic approach,'' \emph{arXiv preprint
  arXiv:1901.07226}, 2019.

\end{thebibliography}
\end{spacing}
\appendices
\section{Mixed Strategy Nash Equilibrium (MSNE)}
\label{sec:appendix_1}
We define $\vect{\tau^{*}} = [\tauD^{*},\tauW^{*}]$ as the parameter required to compute the mixed strategy Nash equilibrium of the one-shot game. We begin by finding the $\tauD^{*}$ of the $\AO$ network by solving the optimization problem 
\begin{equation}
\begin{aligned}
\textbf{OPT I:}\quad& \underset{\tauD}{\text{minimize}}
& & \uD \\
& \text{subject to}
& & 0 \leq \tauD \leq 1.
\end{aligned}
\label{Eq:opt1}
\end{equation}
where, $\uD$ is the payoff of the $\AO$ network defined as
\begin{align*}
\uD & = (1-\tauD(1-\tauD)^{(\ND-1)}(1-\tauW)^{\NW})\AvginitAoIT{}{}\\
&+ (1-\tauD)^{\ND}(1-\tauW)^{\NW}(\lidle-\lcol) + \lcol\\
&+ (\ND\tauD(1-\tauD)^{(\ND-1)}(1-\tauW)^{\NW}\\
&+ \NW\tauW(1-\tauW)^{(\NW-1)}(1-\tauD)^{\ND})(\lsucc-\lcol).
\end{align*}
The Lagrangian of the optimization problem~(\ref{Eq:opt1}) is
\begin{align*}
\mathcal{L}(\tauD,\mu) = & \uD -\mu_{1}\tauD + \mu_{2}(\tauD-1).
\end{align*}
where $\vect{\mu} = [\mu_{1},\mu_{2}]^{T}$ is the Karush-Kuhn-Tucker (KKT) multiplier vector. The first derivative of the objective function in~(\ref{Eq:opt1}) is
\footnotesize
{\begin{align*}
\uD' & = -\AvginitAoIT{}{}(1-\tauW)^{\NW}[(1-\tauD)^{(\ND-1)} - (\ND-1)\nonumber\\
&\hspace{-1.5em}\tauD(1-\tauD)^{(\ND-2)}] + (\lsucc-\lcol)[(1-\tauW)^{\NW}(\ND(1-\tauD)^{(\ND-1)}\nonumber\\
&\hspace{-1.5em}-\ND(\ND-1)\tauD(1-\tauD)^{(\ND-2)})-\ND\NW\tauW(1-\tauW)^{(\NW-1)}\nonumber\\
&\hspace{-1.5em}(1-\tauD)^{\ND-1}]-(\lidle-\lcol)\ND(1-\tauW)^{\NW}(1-\tauD)^{(\ND-1)}.
\end{align*}}
\normalsize
The KKT conditions can be written as
\begin{subequations}
\begin{align}
\uD-\mu_{1} + \mu_{2} &= 0,\label{Eq:stationarity}\\
-\mu_{1}\tauD &= 0,\label{Eq:complementarySlackness_1}\\
\mu_{2}(\tauD-1) &= 0,\label{Eq:complementarySlackness_2}\\
-\tauD &\leq 0,\\
\tauD-1 &\leq 0,\\
\vect{\mu} = [\mu_{1},\mu_{2}]^{T} &\geq 0.\label{Eq:complementarySlackness_3}
\end{align}
\end{subequations}
We consider three cases. In case (i), we consider $\mu_{1} = \mu_{2} = 0$. From the stationarity condition~(\ref{Eq:stationarity}), we get
\begin{align}
\tauD = \frac{(1-\tauW)(\AvginitAoIT{}{}-\ND(\lsucc-\lidle))+\ND\NW\tauW(\lsucc-\lcol)}{\left(\splitfrac{(1-\tauW)\ND(\AvginitAoIT{}{}+(\lidle-\lcol)-\ND(\lsucc-\lcol))}{+\ND\NW\tauW(\lsucc-\lcol)}\right)}.
\label{Eq:tauD}
\end{align}
In case (ii) we consider $\mu_{1} \geq 0, \mu_{2} = 0$. Again, using~(\ref{Eq:stationarity}), we get $\mu_{1} = \uD'$. From~(\ref{Eq:complementarySlackness_3}), we have $\mu_{1}\geq 0$, therefore, $\uD' \geq 0$. On solving this inequality on $\uD'$ we get, $\AvginitAoIT{}{} \leq \AvginitAoIT{}{\text{th},0}$, where $\AvginitAoIT{}{\text{th},0} = \ND(\lsucc - \lidle) - \frac{\ND\NW\tauW(\lsucc-\lcol)}{(1-\tauW)}$.

Finally, in case (iii) we consider $\mu_{1} = 0, \mu_{2} \geq 0$. On solving~(\ref{Eq:stationarity}), we get $\AvginitAoIT{}{} \leq \AvginitAoIT{}{\text{th},1}$, where $\AvginitAoIT{}{\text{th},1} = \ND(\lsucc - \lcol)$.

Therefore, the solution from the KKT condition is
\scriptsize{
\begin{align}
&\tauD^{*} = 
    \begin{cases}
     \hspace{-0.5em}
      \begin{aligned}
 	  \frac{(1-\tauW)(\AvginitAoIT{}{}-\ND(\lsucc-\lidle))+\ND\NW\tauW(\lsucc-\lcol)}{\splitfrac{(1-\tauW)\ND(\AvginitAoIT{}{}+(\lidle-\lcol)-\ND(\lsucc-\lcol))}{+\ND\NW\tauW(\lsucc-\lcol)}}
 	  \end{aligned}
	  \vspace{1em}
      &\hspace{-0.75em}
      \begin{aligned}
      &\AvginitAoIT{}{}> \AvginitAoIT{}{\text{th}},
      \end{aligned}\\
      1&\hspace{-8em}
      \begin{aligned}
      \AvginitAoIT{}{}< \AvginitAoIT{}{\text{th}}\text{ \& }\AvginitAoIT{}{\text{th}} = \AvginitAoIT{}{\text{th},1},
      \end{aligned}\\
      0&\hspace{-8em}
      \begin{aligned}
      \AvginitAoIT{}{}< \AvginitAoIT{}{\text{th}}\text{ \& }\AvginitAoIT{}{\text{th}} = \AvginitAoIT{}{\text{th},0}.
      \end{aligned}
    \end{cases}
\label{Eq:KKT_OPT1_temp}
\end{align}}
\normalsize
where, $\AvginitAoIT{}{\text{th}} = \max\{\AvginitAoIT{}{\text{th},0},\AvginitAoIT{}{\text{th},1}\}$. Under the assumption that length of successful transmission is equal to the length of collision i.e. $\lsucc = \lcol$,~(\ref{Eq:KKT_OPT1_temp}) reduces to
\begin{align}
\tauD^{*} = 
\begin{cases}
  \begin{aligned}
  \frac{\ND(\lidle-\lsucc)+\AvginitAoIT{}{}}{\ND(\lidle-\lcol+\AvginitAoIT{}{})}
  \end{aligned}
  &
  \begin{aligned}
  &\AvginitAoIT{}{}>\ND(\lsucc-\lidle),
  \end{aligned}\\
  0 &\text{ otherwise }.
\end{cases}
\label{Eq:KKT_OPT1}
\end{align}

Similarly, we find $\tauW^{*}$ for the $\TO$ network by solving the optimization problem
\begin{equation}
\begin{aligned}\textbf{OPT II:}\quad& \underset{\tauW}{\text{minimize}}
& & -\uW \\
& \text{subject to}
& & 0 \leq \tauW \leq 1.
\end{aligned}
\label{Eq:opt2}
\end{equation}
where, $\uW$ is the payoff of the $\TO$ network defined as
\begin{align*}
\uW & = \tauW(1-\tauW)^{(\NW-1)}(1-\tauD)^{\ND}\lsucc.
\end{align*}
The Lagrangian of the optimization problem~(\ref{Eq:opt2}) is
\begin{align*}
\mathcal{L}(\tauW,\mu) = & -\uW -\mu_{1}\tauW + \mu_{2}(\tauW-1).
\end{align*}
where $\vect{\mu} = [\mu_{1},\mu_{2}]^{T}$ is the KKT multiplier vector. The first derivative of $\uW$ is
\begin{align*}
\uW' &= (1-\tauD)^{\ND}(1-\tauW)^{(\NW-1)}\lsucc\\
& - (\NW-1)\tauW(1-\tauW)^{(\NW-2)}(1-\tauD)^{\ND}\lsucc.
\end{align*}
The KKT conditions can be written as
\begin{subequations}
\begin{align}
-\uW' -\mu_{1} + \mu_{2} &= 0,\label{Eq:stationarityWiFi}\\
-\mu_{1}\tauW &= 0,\label{Eq:complementarySlacknessWiFi_1}\\
\mu_{2}(\tauW-1) &= 0,\label{Eq:complementarySlacknessWiFi_2}\\
-\tauW &\leq 0,\\
\tauW-1 &\leq 0,\\
\vect{\mu} = [\mu_{1},\mu_{2}]^{T} &\geq 0.\label{Eq:complementarySlacknessWiFi_3}
\end{align}
\end{subequations}
We consider the case when $\mu_{1} = \mu_{2} = 0$. From the~(\ref{Eq:stationarityWiFi}), we get $\uW' = 0$. On solving the stationarity condition, we get $\tauW^{*} = 1/\NW$, which is also the solution of the KKT conditions.
\end{document}